# EXPERIENCE WITH THE PEP-II RF SYSTEM AT HIGH BEAM CURRENTS*

P. Corredoura[†], S. Allison, W. Ross, R. Sass, R. Tighe
Stanford Linear Accelerator Center, Stanford, Ca 94309, USA

*Abstract*

The PEP-II B Factory Low-Level RF System (LLRF) is a fully programmable VXI based design running under an EPICS control environment. Several RF feedback loops are used to control longitudinal coupled-bunch modes driven by the accelerating mode of the RF cavities. This paper updates the performance of the LLRF system as beam currents reach design levels. Modifications which enhance the stability, diagnostics, and overall operational performance are described. Recent data from high current operation is included.

## 1. INTRODUCTION

Both the high energy ring (HER) and the low energy ring (LER) of the PEP-II B factory are inherently longitudinally unstable due to interaction between the beam and the fundamental mode of the RF cavities [1]. RF feedback loops operating at baseband and a fiber optic connection to the longitudinal multibunch feedback system were used successfully to control the low-order longitudinal modes (figure 3) [2,3,4]. The system is modular and very compact. The LLRF hardware for each RF station is based on a set of custom, highly integrated VXI modules (figure 1). Each station contains a built-in network analyzer to configure and test the feedback loops and a series of transient signal recorders which can record a wide variety of waveforms. After a fault has occurred the circular buffers are frozen and the data is stored in files. This post-mortem analysis capability has proven to be extremely beneficial for diagnosing problems, especially intermittent faults.

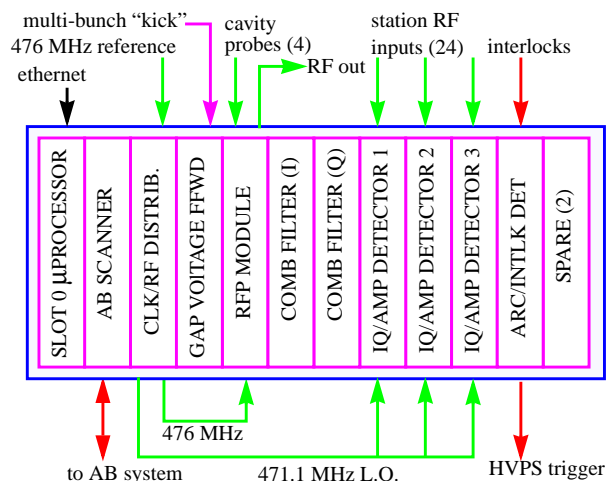

Fig. 1. PEP-II LLRF system VXI crate topology (HER)

*Work supported by Department of Energy, contract DE-AC03-76SF00515

[†] plc@slac.stanford.edu

Overall the system has preformed extremely well. To date stored current has reached 950 mA in the HER and 1700 mA in the LER. Even at the highest currents the beams are well stabilized by the active controls. Presently the HER beam current is limited to the 700 mA due to heating of one vacuum chamber. Beam current and phase of both rings while colliding are shown (figure 2).

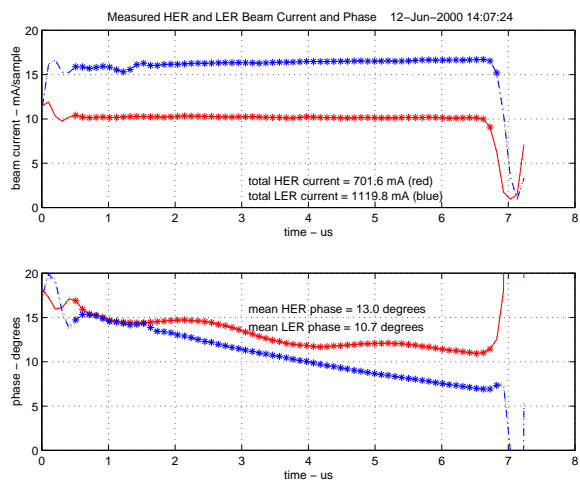

Fig. 2. Beam current and phase profiles measured during colliding with hardware built into the LLRF system. The sinusoidal shaped HER phase is due to four parked cavities.

The use of RF feedbacks does complicate some aspects of system operation. The changes we have made and plan to make to improve operation will be described in detail so others may benefit from our experience.

## 2. DECREASING STATION CYCLE TIME

Once beam currents exceed ~300 mA, a beam abort will cause all the RF stations to trip on cavity reflected power. This is due to the cavities tuners being set to match the large reactive beam contribution, which is suddenly removed. In addition, the wideband direct RF feedback loop attempts to maintain the gap voltage (figure 3), keeping output power high while cavities are reflecting >90kW of RF power. Since the tuners cannot move quickly we chose to improve the overall station power-up cycle time.

The power-up procedure is implemented as an EPICS state sequence, thus is fully programmable [5]. The original approach (slow turn-on) energized the station at a moderate gap voltage with RF feedback loops disabled. The gain of the direct and then the comb loop were ramped up, followed by raising the gap voltage to the desired level. This procedure required three minutes to complete. The new fast turn-on scheme presets the tuner positions, loop gains, and baseband IQ references to their no-beam, normal gap voltage values. The klystron high voltage corresponding to the no-beam condition is then applied. Once the klystron reaches nominal output all the feedback loops settle into regular operation. This technique reduced the cycle time to less than 20 seconds, but there were complications...

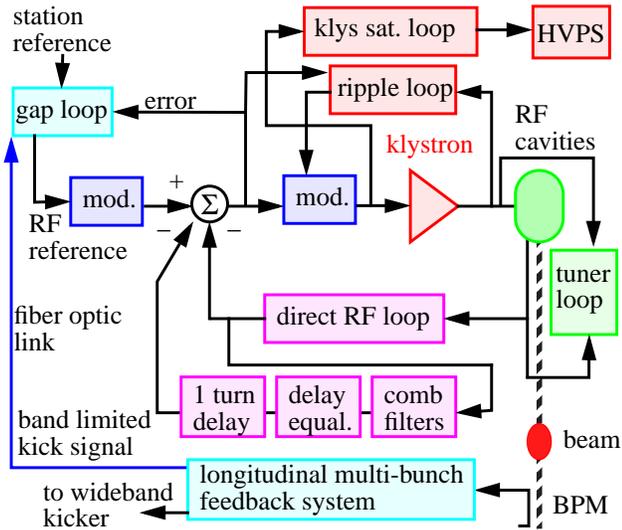

Fig. 3. Block diagram of RF feedback loops used in the PEP-II low-level RF system. Multi-cavities not shown.

## 3. EXCESSIVE BASEBAND VOLTAGES

When direct RF feedback is employed, the drive power is proportional to the direct loop error voltage. In an IQ baseband RF system like PEP-II, the error signals are voltages representing the real and imaginary components of what eventually will become the klystron RF drive. During operational conditions when the error signals become large, excessive baseband voltages may present a problem.

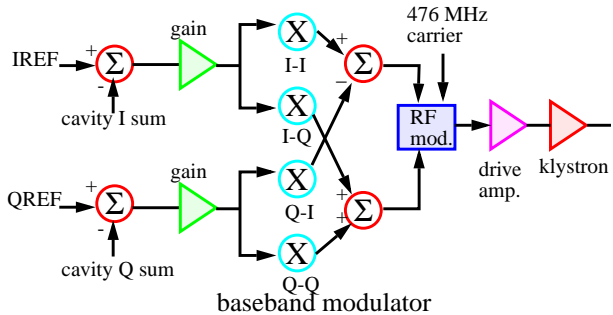

Fig. 4. PEP-II LLRF baseband drive analog processing.

The analog signal chain for the PEP-II LLRF system is depicted in figure 4. The baseband modulator is composed of four wideband analog multipliers followed by a pair of high speed op-amps [4]. This particular modulator is used to compensate for klystron gain and phase variations due to changes in the cathode voltage as power requirements vary (PEP-II klystrons do not contain modulation anodes). The multipliers are Gilbert-cell based devices rated at 1 volt maximum input. An idiosyncrasy of this class of multipliers is that when over-driven the output polarity may actually invert. This would result in an incorrect phase shift across the modulator and may cause positive feedback around the direct RF feedback loop. This effect complicated our initial attempts for fast turn-on.

To prevent overdriving the multipliers, back-to-back Schottky diodes were installed across the feedback resistor in the gain stage preceding the baseband modulator. A series resistor was added to form a "soft" limiter at the desired 1 volt threshold (figure 5). The addition of the limiting circuit greatly improved the performance of the fast turn-on approach and is presently in use. There is, however, a further complication...

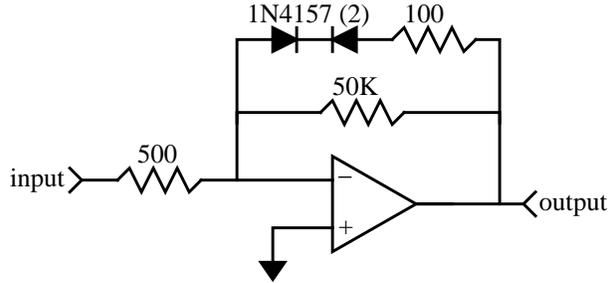

Fig. 5. Schematic of gain amplifier with limiting diodes.

## 4. SECONDARY EFFECT OF KLYSTRON GAIN TRACKING

As klystron cathode voltage is elevated to increase output power, the gain of the tube increases. This effect is most dramatic in the LER since each klystron drives two rather then four cavities. The required klystron power ranges from 250 kW (no beam) up to the rated 1.2 MW output, an increase of 7dB. Since the klystron is in the direct RF feedback loop, the gain of the baseband modulator must be decreased to compensate for the additional klystron gain. Decreasing the modulator gain forces an increase in the multiplier input voltages in order to keep drive power constant. This implies that enough dynamic range must be available to allow the multiplier input voltages to double and still remain less than 1 volt.

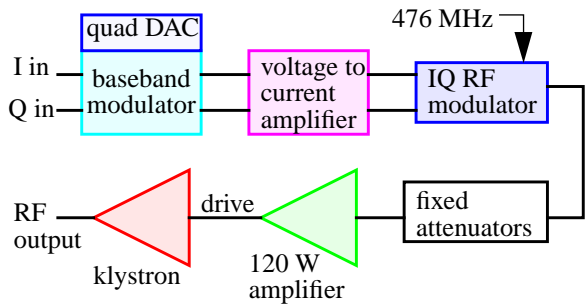

Fig. 6. Block diagram showing components responsible for determining how IQ voltages scale to drive power.

Configuring the drive chain (figure 6) such that at the maximum modulator gain (corresponding to the no-beam klystron power) the IQ voltages are half scale is achieved by picking the proper transimpedance gain and fixed attenuators. During conditions when the IQ voltages are saturated (during fast turn-on for example), the drive power could exceed four times the desired value. We avoid this during fast turn-on by presetting the baseband modulator gain to the minimum allowed (full beam current) setting.

## 5. PREVENTING EXCESSIVE DRIVE POWER FOR ROBUST OPERATION

During normal operation we have experienced occasional, as yet unexplained, sudden loss of the cavity probe RF signal. The built-in transient recorders have proven very useful to study these phenomenon. The fault signature is not consistent with a full cavity arc since the reflected power does not show a large perturbation. In systems operating with direct RF feedback the loss of a cavity probe signal will cause the drive chain to immediately saturate. If the klystron is severely over-driving, the output power will actually drop (figure 7). If the drive power exceeds the peak of the saturation curve, even for a brief instant, the direct feedback loop will saturate and not recover.

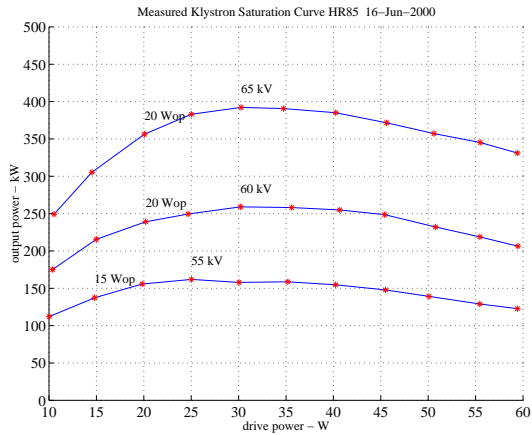

Fig. 7. Measured klystron saturation curve showing suggested operating points (Wop) at various cathode voltages.

In an attempt to prevent klystron overdrive we configured the transimpedance amplifier gain such that the RF modulator was operating near saturation when producing saturated klystron drive power from one baseband input (I or Q). This implies that the maximum power could be 3dB higher if both inputs were driven to saturation.

In practice the task of balancing maximum available drive power with required dynamic range of the baseband modulator is a trade off. Operating any of the components near saturation degrades the system's ability to amplitude modulate. The ability of the LLRF system to fully cancel the RF ripple caused by the switching aspect of the high voltage power supply (HVPS) is compromised. Based on our experience we propose upgrading the system.

## 6. PROPOSED CHANGES

By applying active limiters before the baseband modulator we can prevent overdriving the multipliers while preserving linearity during normal operation (figure 8). Suitable op-amps must be selected which can tolerate the +/-1 volt differential voltages present during non-limiting condition.

A second limiting circuit will be added to prevent the drive power from exceeding the klystron's saturation point (figure 9). The actual drive power will be detected by an existing linear detector in the IQA module [6]. If the detected voltage exceeds a programmable set-point, both the baseband drive signals will be reduced proportionally to decrease the drive power to a programmable level while maintaining the output phase. We expect this feature will allow each RF system to "ride through" transient events which presently cause a fault.

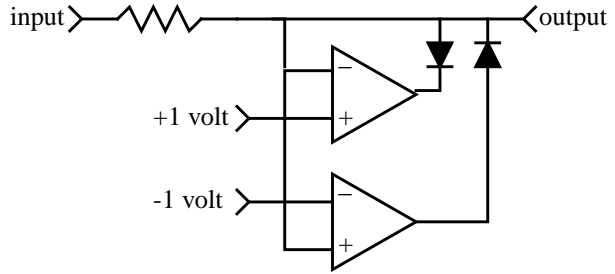

Fig. 8. Bipolar limiter to prevent overdrive of multipliers.

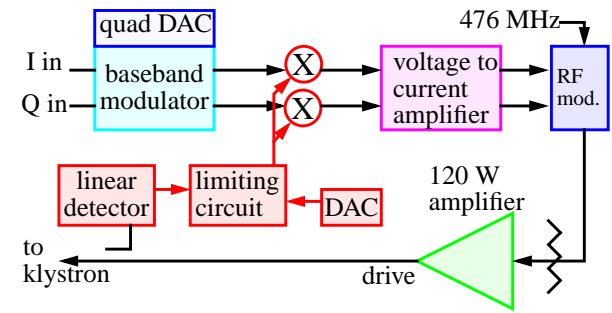

Fig. 9. Drive power limiting circuit to improve robustness.

The final change we plan to make to the system is the addition of a wideband analog "ripple" loop to cancel RF modulation caused by the switching aspect of the HVPS. A DSP was intended to handle this task but the combination of significant delay in the digital IQ receiver and the 50 kHz bandwidth ripple proved challenging. Presently an analog integrator in the direct RF feedback loop cancels the ripple but simulations show it will cause instability as beam currents reach 2A.

## 7. CONCLUSIONS

The PEP-II RF system has performed extremely well. Large beam currents are routinely stored with no sign of instabilities. The EPICS interface provides a flexible tool to improve many operational aspects of the system in short order. We hope our experience with baseband RF feedbacks will provide others with useful information.